\documentclass[pre,twocolumn,psfig]{revtex4}
\usepackage{graphicx,amssymb}
\input psfig.sty

\begin{document}

\title{Preferential attachment growth model and nonextensive statistical mechanics}

\author{Danyel J.B. Soares$^{1}$, Constantino Tsallis$^{2, 3}$, Ananias M. Mariz$^{1}$
and Luciano R. da Silva$^{1}$}
\thanks{E-mail addresses:
 danyel@dfte.ufrn.br, tsallis@cbpf.br, ananias@dfte.ufrn.br, luciano@dfte.ufrn.br
} 
\address{
$^1$Departamento de F\'{\i}sica Te\'{o}rica e Experimental,
Universidade Federal do Rio Grande do Norte \\ Campus
Universitario, 59072-970 Natal-RN, Brazil\\ $^2$Centro Brasileiro
de Pesquisas F\'{\i}sicas, Rua Xavier Sigaud 150, 22290-180 Rio de
Janeiro-RJ, Brazil.\\ $^3$Santa Fe Institute, 1399 Hyde Park Road,
Santa Fe, New Mexico 87501,  USA. }

\date{\today}

\begin{abstract}
We introduce a two-dimensional growth model where every new site is located, at a
distance $r$ from the barycenter of the pre-existing graph, according to the
probability law $1/r^{2+\alpha_G} \;(\alpha_G \ge 0)$, and is attached to (only) one
pre-existing site with a probability  $\propto k_i/r^{\alpha_A}_i \;(\alpha_A \ge 0$;
$k_i$ is the number of links of the $i^{th}$ site of the pre-existing graph, and $r_i$
its distance to the new site).  Then we numerically determine that the probability
distribution for a site to have $k$ links is asymptotically given, for all values of
$\alpha_G$, by $P(k) \propto e_q^{-k/\kappa}$, where $e_q^x \equiv
[1+(1-q)x]^{1/(1-q)}$ is the function naturally emerging within nonextensive
statistical mechanics. The entropic index is numerically given (at least for $\alpha_A$ not too large) 
by $q = 1+(1/3)
e^{-0.526 \; \alpha_A}$, and the characteristic number of links  by $\kappa \simeq
0.1+0.08 \, \alpha_A$. The $\alpha_A=0$ particular case belongs to the same
universality class to which the Barabasi-Albert model belongs. In addition to this, we
have numerically 
studied the rate at which the average number of links $\langle k_i \rangle$
increases with the scaled time $t/i$; asymptotically,   $\langle k_i \rangle \propto
(t/i)^\beta$, the exponent being close to $\beta=\frac{1}{2}(1-\alpha_A)$ for $0 \le
\alpha_A \le 1$, and zero otherwise.
 The present results reinforce the conjecture that the microscopic dynamics of
nonextensive systems typically build (for instance, in Gibbs $\Gamma$-space for Hamiltonian systems) a
scale-free network.
\end{abstract}
\maketitle

Among the subjects that are being studied intensively nowadays in statistical physics,
there are two, namely  {\it nonextensive statistical mechanics} (see
\cite{gellmanntsallis} for a review) and
{\it networks} \cite{watts}, in particular
{\it scale-free} networks \cite{barabasi1}, which receive special attention in
connection with {\it complex systems} \cite{gellmann,barabasi2,barabasivarious,yookmanna}. Could
these two topics be intimately
related? This would not be so surprising after all, given the fact that both research
lines frequently address  similar types of natural and artificial systems, in physics,
economics, chemistry, biology, linguistics, social sciences and others. In fact, such a
connection has already been  conjectured in several occasions, e.g.,
\cite{gellmanntsallis} (Preface and Chapter 1) and \cite{conjecture}. In the present
paper we propose a growth model, on which we exhibit and quantitatively explore this
connection.

Let us consider a continuous plane. We shall construct a single connected network of
{\it sites} (or {\it nodes} or {\it vertices}) and {\it links} (or {\it bonds} or {\it
edges}) by gradually (sequentially) making it grow. We first fix one site ($i=1$) at
some arbitrary origin of the plane. The second site ($i=2$) is randomly and
isotropically chosen at a distance $r$ distributed according to the probability law
$P_G(r) \propto 1/r^{2+\alpha_G} \;(\alpha_G \ge 0$; $G$ stands for {\it growth}). This
second site is then linked to the first one. To locate the third site ($i=3$) we move
the origin to the barycenter of the two first sites, and apply again the distribution
$P_G(r)$ from this new origin. This third site is now going to be linked to only one of
the pre-existing two sites. To do this, we use an attachement probability $p_A \propto
k_i/r^{\alpha_A}_i \;(\alpha_A \ge 0$; $A$ stands for {\it attachment}), where $r_i$ is
the distance of the newly arrived site to the $i^{th}$ site of the pre-existing
cluster, and the {\it connectivity} $k_i$ is the number of links already arriving to
the same $i^{th}$ site (at the present stage, $k_1=k_2=1$). The growth-attachment
process is sequentially repeated as long as we want; the sites have unit radius, and no
new site is admitted which overlaps with a previous one. If we denote with $N$ the
total number of sites of the  cluster, it immediately follows that the linking of the
newly arrived site ($i=N$) is done with the probability
\begin{equation}
p_A = \frac{k_i/r^{\alpha_A}_i}{\sum_{j=1}^{N-1} k_j/r^{\alpha_A}_j}
\end{equation}

\begin{figure}
\vspace{.2in}
\includegraphics[height=6.0cm,width=6.1cm,angle=0]{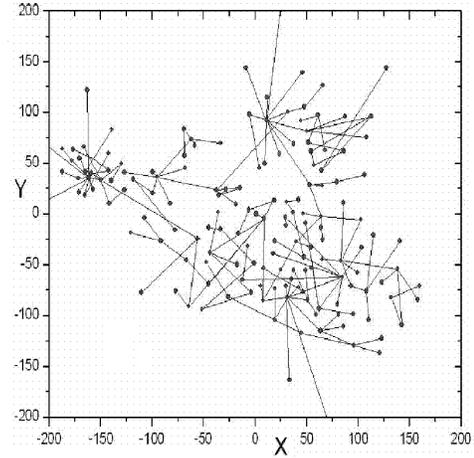}
\caption{\small Typical $N=250$ network for
$(\alpha_G,\alpha_A)=(1,0)$. The starting site at $(X,Y)=(0,0)$ is
indicated with a larger circle. Notice the spontaneous emergence
of hubs. \label{fig1}}
\end{figure}


It is clear that the dynamics of this model makes the arriving sites to have
preferential attachment to the previous sites that already have many links ({\it
hubs}), {\it as long as they are not too far}. This competition (already explored in \cite{yookmanna} for the particular case of {\it uniform} distribution of sites within some limited region) between connectivity
and (Euclidean) proximity is less pronounced when $\alpha_A$ is close to zero, and completely
disappears only at $\alpha_A=0$. For this particular case, one expects behaviors
consistent with the Barabasi-Albert model \cite{barabasi2}, which has topology but no
metrics. In the present paper, we focus on two main aspects: (i) the stationary-state
connectivity distribution $P(k)$ associated with the number of sites that have $k$
links in the $N \to\infty$ limit; (ii)  the time dependence of the average number
$\langle k_i \rangle$ of links, more precisely how $\langle k_i \rangle$ grows with the
scaled time $t/i$ ($t \ge i$), particularly in the limit $t/i \to \infty$ (see
\cite{barabasi2}).

\begin{figure}
\vspace{.2in}
\resizebox{70mm}{!}{\includegraphics[width=\columnwidth,angle=0]{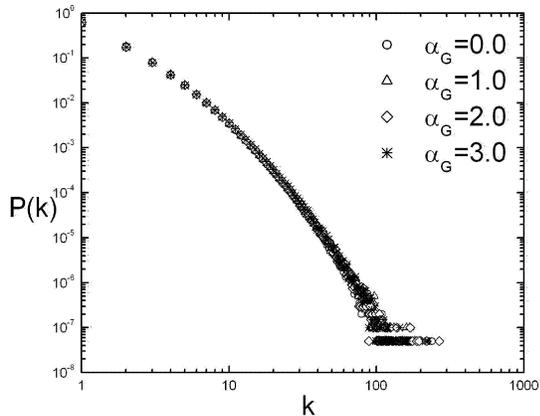}}
\caption{\small Connectivity distribution for
$\alpha_A=1$ and typical values of $\alpha_G$; 2000 realizations
of $N=10\:000$ networks .} \label{fig2}
\end{figure}

Typical networks obtained with this model are indicated in Fig. 1. Our numerical results
for $P(k)$ are indicated in Figs. 2 and 3. We illustrate in Fig. 2 the fact that $P(k)$
does not depend from $\alpha_G$ for any given value of $\alpha_A$. The exponent
$\alpha_G$ controls the metrics of the emerging cluster, but has no influence on the
connectivity distribution. In contrast, this distribution is greatly influenced by the
exponent $\alpha_A$, as illustrated in Fig. 3.
We can verify that all our examples are very well fitted with the form
\begin{equation}
P(k) = P(0) \,e_q^{-k/\kappa}\;,
\end{equation}
where the {\it $q$-exponential} function is defined as follows
\cite{quimicanova,gellmanntsallis}
\begin{equation}
e_q^x \equiv [1+(1-q)\,x]^{1/(1-q)} \;\;\;(e_1^x=e^x) \; ;
\end{equation}
$\kappa>0$ is a characteristic number of links. In Fig. 4 we show the $\alpha_A$
dependences of $q$ and $\kappa$. In particular, $q(\alpha_A)$ does not exhibit a critical value of $\alpha_A$ above which a different regime could emerge. However, unless a detailed analysis (out of the aim of the present letter) is done of the finite-size effects, such a possibility should not be excluded.

\begin{figure}
\vspace{.2in}
\begin{tabular}{c}
\resizebox{70mm}{!}{\includegraphics[width=\columnwidth,angle=0]{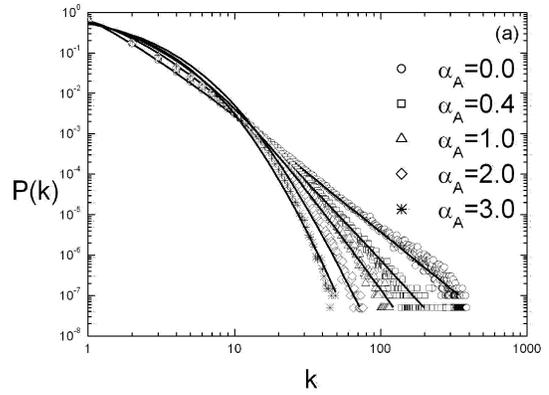}}\\[2.5cm]
\resizebox{70mm}{!}{\includegraphics[width=\columnwidth,angle=0]{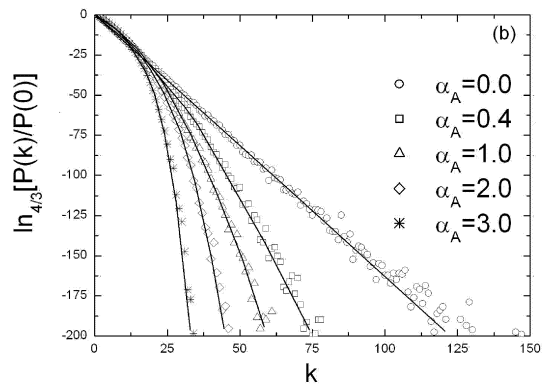}}\\[2.5cm]
\resizebox{70mm}{!}{\includegraphics[width=\columnwidth,angle=0]{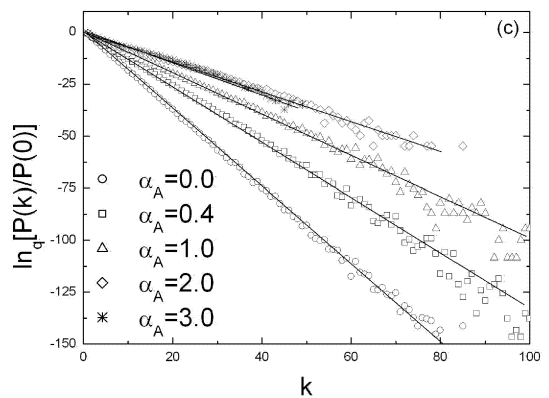}}
\end{tabular}
\caption{\small
Connectivity distribution for typical values of $\alpha_A$ (we have used $\alpha_G=2$
but we recall that this value is irrelevant). Points are our computer simulation
results; continuous lines are the best fits with the $q$-exponential function indicated
in Eq. (2). (a) $log-log$ representation; (b) $\ln_{4/3}-linear$ representation, with
$\ln_q x \equiv \frac{x^{1-q}-1}{1-q}$; (c)  $\ln_{q}-linear$ representation, where,
for each value of $\alpha_A$, we have used its corresponding value of $q$. We have used
3 different representations to improve comprehension.}
\label{fig3}
\end{figure}

\begin{figure}
\vspace{.2in}
\begin{tabular}{c}
\resizebox{70mm}{!}{\includegraphics[width=\columnwidth,angle=0]{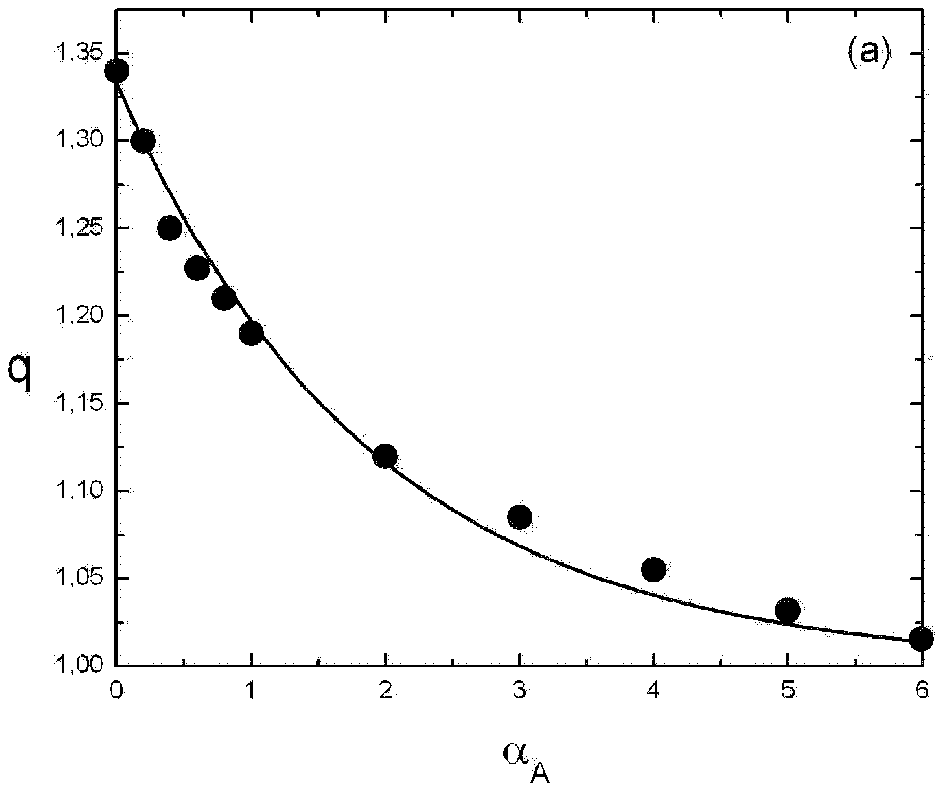}}\\[2.5cm]
\resizebox{70mm}{!}{\includegraphics[width=\columnwidth,angle=0]{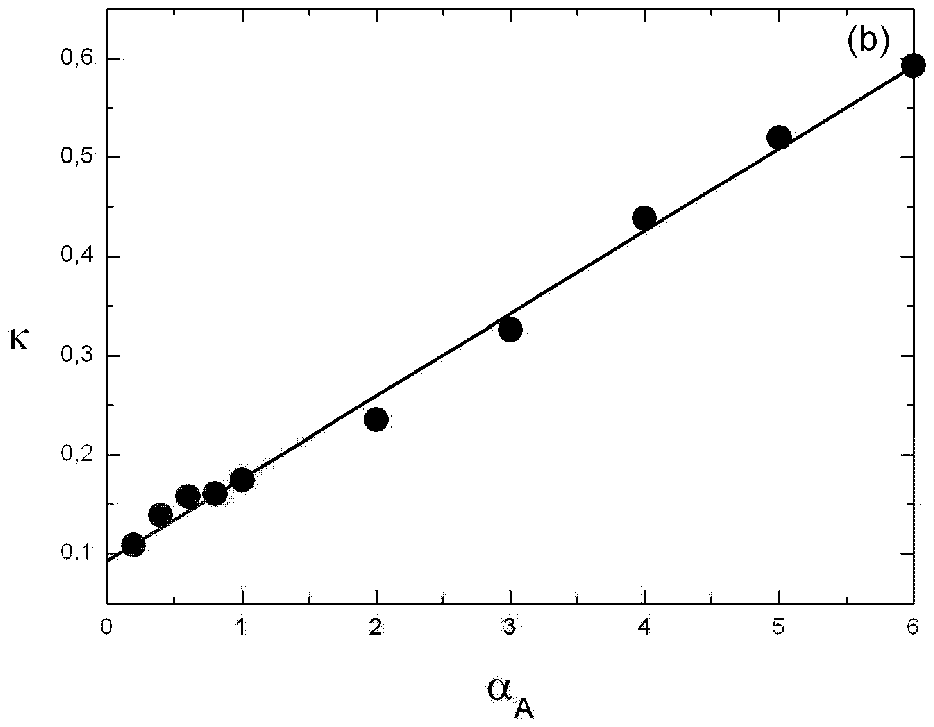}}
\end{tabular}
 \caption{\small Values of $q$ and $\kappa$ used in
the best fits indicated in Fig. 3. The solid curves are: (a) $q =
1+(1/3) e^{-0.526 \; \alpha_A} \;(\forall \;\alpha_G)$; (b)
$\kappa \simeq 0.083+0.092 \, \alpha_A \;(\forall \;\alpha_G)$.}
\label{fig4}
\end{figure}


\begin{figure}
\vspace{.2in}
\begin{tabular}{c}
\resizebox{70mm}{!}{\includegraphics[width=\columnwidth,angle=0]{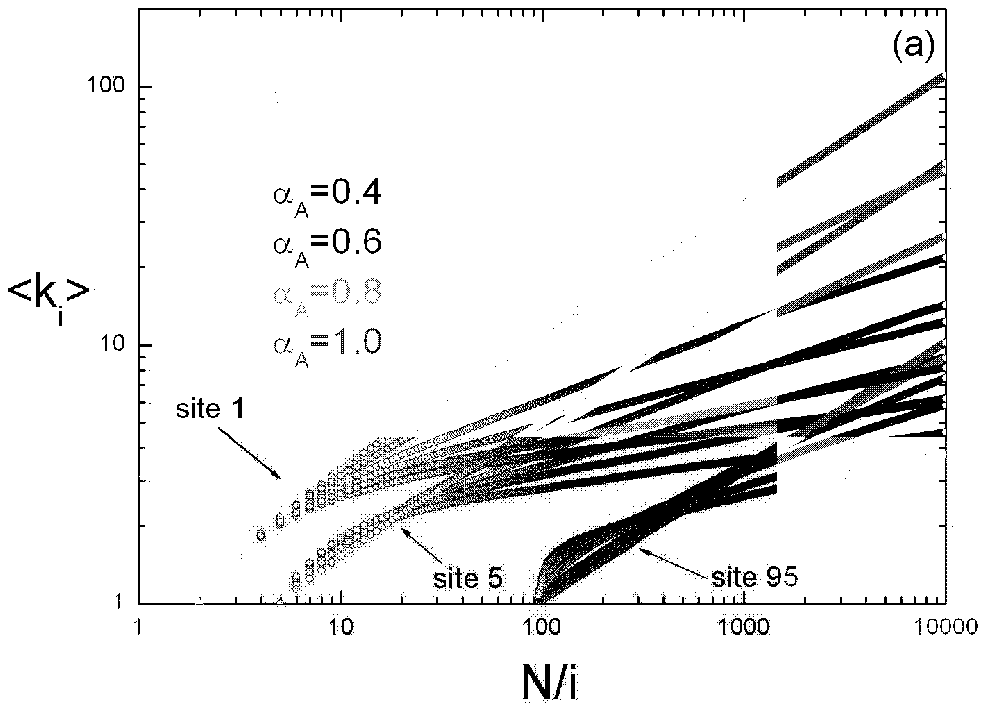}}\\[2.5cm]
\resizebox{70mm}{!}{\includegraphics[width=\columnwidth,angle=0]{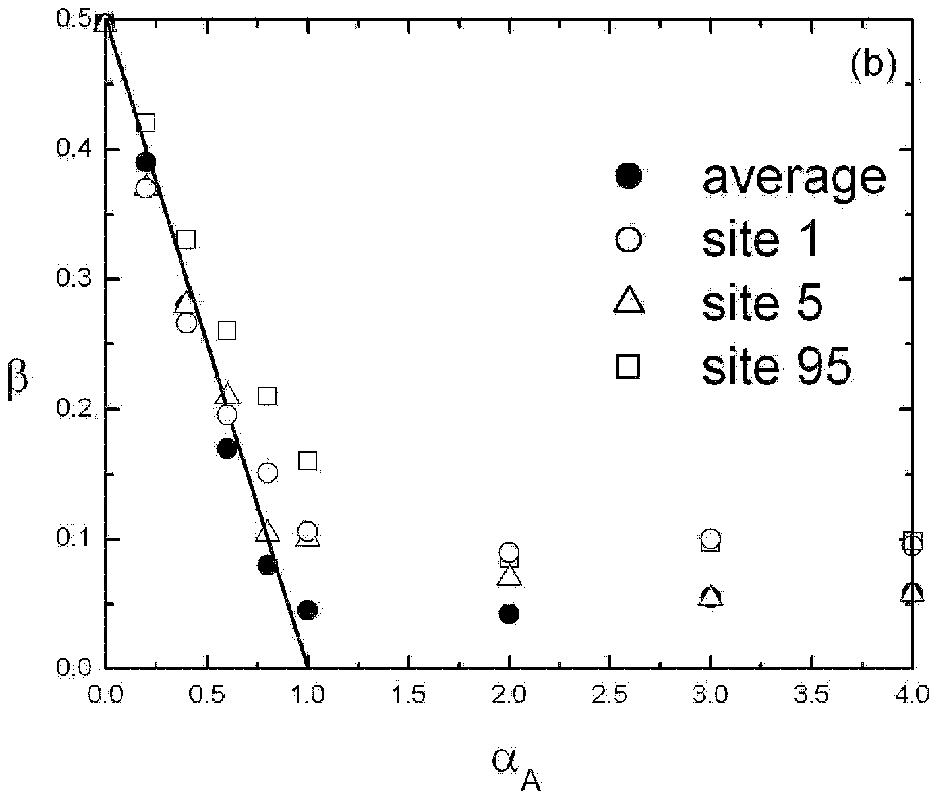}}
\end{tabular}
\caption{(a) Time dependence ($t=N$) of the average number (over 2000 realizations) of links for
typical values of $\alpha_A$ for sites $i=1$, $i=5$ and $i=95$ (compare with Fig. 2(c)
of \cite{barabasi2}). We have used $\alpha_G=2$; (b) $\alpha_A$-dependance of $\beta$
(the straight line $\beta= \frac{1}{2}(1-\alpha_A)$ could be the exact answer).
}
\label{fig5}
\end{figure}

We must now address a relevant point: Why have we fitted our curves with
$q$-exponentials? The reason lies on the conjectured connection with nonextensive
statistical mechanics \cite{tsallis88} (see \cite{gellmanntsallis,tsallisreviews} for
reviews), a theory which we briefly illustrate now.
Let us consider the following generalization of the Boltzmann-Gibbs entropy
$S_{BG}=-\int dk\, P(k) \ln P(k)$:
\begin{equation}
S_q=\frac{1-\int dk \,[P(k)]^q}{q-1} \;\;\;\;(q \in {\cal R};\,S_1=S_{BG})
\end{equation}
If we optimize this entropy with the constraints $\int dk \,P(k)=1$ and $\{\int dk
\,k\,[P(k)]^q\} / \{   \int dk \,[P(k)]^q\} =K$ we obtain straightforwardly \cite{TMP}
\begin{equation}
P(k)=\frac{e_q^{-\beta_q(k-K)}}{\int dk^\prime\, e_q^{-\beta_q(k^\prime-K)}} \; ,
\end{equation}
where $\beta_q$ is simply related with a Lagrange parameter. This expression precisely
coincides with Eq. (2) through the identification
\begin{equation}
P(0) \equiv \frac{  e_q^{\beta_q K}}   {  \int dk^\prime\, e_q^{-\beta_q(k^\prime-K)} }
\;,
\end{equation}
\begin{equation}
\kappa \equiv \frac{1+(1-q) \beta_q K}{\beta_q}
\end{equation}

Before proceeding with our numerical results, let us briefly mention what indications
make us to believe that a close connection, and not just a mere functional coincidence,
might exist between the present preferential-attachement growth model and the
thermostatistical systems addressed by the entropy (4). First, a growth model
involving, like the present one, preferential attachment has been proposed and both
analytically and numerically discussed in \cite{barabasi3}. In this model the
connectivity distribution is analytically shown to be precisely of the form (2)
(although written in a slightly different manner). The corresponding entropic index is
given \cite{gellmanntsallis} (Chapter 1) by $q = [2m(2-r)+1-p-r]/[m(3-2r)+1-p-r]$,
where $(m,p,r)$ are parameters of the model.
Second, Lennard-Jones small clusters (with up to 14 atoms) have been numerically studied
recently \cite{doye}. The distributions of the number of local minima of the potential
energy  with $k$ neighboring saddle-points in the configurational phase space can,
although not mentioned in the original paper \cite{doye}, be quite well fitted with
$q$-exponentials with $q=2$ \cite{conjecture}.
Third, the present model generates structures that are scale-free (see, for instance,
Fig. 1). Consistently, there is plethoric evidence of the connections of the
nonextensive entropy $S_q$, and of its associated statistics, with hierarchical and
(multi)fractal structures (see, for instance,
\cite{fractals,fractals2,fractals3,fractals4}). It is clear that none of these three
features constitutes a proof; however, the set of them suggests the quite plausible
scenario that the present growth model basically satisfies the dynamical requirements
for the nonextensive concepts to be applicable.

Let us now present our results concerning the rate at which the
number of links of a given site increases with time. A basic
quantity is the average (over 2000 realizations) number $\langle
k_i \rangle$ of links at a given time, more precisely as a
function of the {\it scaled time} $t/i$ \cite{barabasi2}. Our
results are illustrated in Fig. 5. 
We numerically find that, for all values of $(\alpha_A,\alpha_G)$,    $\langle k_i \rangle \propto (t/i)^{\beta(\alpha_A)}$, with $\beta(0)=1/2$.  

Let us summarize the present paper. We have introduced a growth model which has both
preferential attachment and metrics. Every new site of the cluster has to ``decide" to
which one of the pre-existing sites will link itself. In this stochastic choice, there
might be competition between the ``popular" sites (those that already have many links)
and the nearby sites. This competition is more pronounced for increasing value of the
exponent $\alpha_A$; it disppears for $\alpha_A=0$. The connectivity distribution and
the rate of increase of links with time are substantially influenced by $\alpha_A$. The
$\alpha_A=0$ model belongs to the same universality class to which the well known
Barabasi-Albert model belongs. In addition to these results, we have shown that the
connectivity distribution is (numerically) given by the $q$-exponential function that
emerges naturally in the frame of nonextensive statistical mechanics. This fact
supports the conjecture \cite{gellmanntsallis} that the typical occupation of the
accessible phase space of many nonextensive dynamical systems might be, in relevant
stationary states, scale-free. This is in notorious variance with the occupation of
standard, extensive, isolated dynamical systems, which tends to be uniform, and whose
equilibrium state is that prescribed within Boltzmann-Gibbs statistical mechanics.

\section*{Acknowledgments}

We acknowledge useful remarks from A.-L. Barabasi, S. Havlin, S.S. Manna, L.G. Moyano and P. C. da
Silva, as well as partial financial support by Pronex/MCT, Faperj,
Capes and CNPq (Brazilian agencies).

\end{document}